\documentclass[pra,twocolumn,aps,reprint,showpacs,superscriptaddress,groupedaddress]{revtex4}
\usepackage{epsfig,graphicx,tabularx}
\usepackage{amsmath,amssymb,mathrsfs}
\usepackage{physics}
\usepackage{psfrag}
\usepackage{enumitem,bm,latexsym}
\usepackage{verbatim,comment}
\usepackage[normalem]{ulem}
\usepackage[usenames,dvipsnames]{color}
\usepackage[dvipdfmx,colorlinks=true,linkcolor=blue,urlcolor=blue,citecolor=blue]{hyperref}
\newcommand{\beq}{\begin{equation}}
\newcommand{\eeq}{\end{equation}}
\newcommand{\bal}{\begin{aligned}[b]}
\newcommand{\eal}{\end{aligned}}
\newcommand{\beqa}{\begin{eqnarray}}
\newcommand{\eeqa}{\end{eqnarray}}
\newcommand{\im}{i}
\newcommand{\e}{e}
\newcommand{\J}{\mathrm{J}}
\newcommand{\R}{\mathrm{R}}

\begin{document}

\title{Quantum effective action for the bosonic Josephson junction}

\author{K. Furutani}
\affiliation{Dipartimento di Fisica e Astronomia ``Galileo Galilei'', 
Universit\`a di Padova, via Marzolo 8, 35131 Padova, Italy}
\affiliation{Istituto Nazionale di Fisica Nucleare, Sezione di Padova, 
via Marzolo 8, 35131 Padova, Italy}
\author{J. Tempere}
\affiliation{Department of Physics, Universiteit Antwerpen,
Universiteitsplein 1, 2610 Antwerpen, Belgium}
\author{L. Salasnich}
\affiliation{Dipartimento di Fisica e Astronomia ``Galileo Galilei'', 
Universit\`a di Padova, via Marzolo 8, 35131 Padova, Italy}
\affiliation{Istituto Nazionale di Fisica Nucleare, Sezione di Padova, 
via Marzolo 8, 35131 Padova, Italy}
\affiliation{Istituto Nazionale di Ottica del Consiglio Nazionale delle Ricerche, 
via Carrara 2, 50019 Sesto Fiorentino, Italy}

\begin{abstract}
We investigate a bosonic Josephson junction by using 
the path-integral formalism with relative phase and population 
imbalance as dynamical variables. 
We derive an effective only-phase action performing functional integration 
over the population imbalance. 
We then analyze the quantum effective only-phase action, which formally 
contains all the quantum corrections. 
To the second order in the derivative expansion and to the lowest order 
in $\hbar$, we obtain the quantum correction to the 
Josephson frequency of oscillation. 
Finally, the same quantum correction is found by adopting an alternative approach. 
Our predictions are a useful theoretical tool for experiments with atomic or superconducting Josephson junctions. 
\end{abstract}



\maketitle

\section{Introduction}

Two superconductors or superfluids separated by a tunneling barrier 
give rise to the so-called Josephson junction 
\cite{josephson1962,barone1982,vari2017}. 
In contrast to superconducting Josephson junctions, it is possible 
to have a huge population imbalance with atomic Josephson junctions 
due to the appearance of the self-trapping phenomena \cite{smerzi1997}. 
The phase model \cite{leggett1991} is often used to describe the 
quantum behavior of Josephson junctions. This model is based on the 
quantum commutation rule between the number-difference operator and 
the phase-difference operator \cite{luis1993}. 
Because the phase-number commutation rule is approximately 
correct for systems with a large number of condensed electronic 
Cooper pairs or bosonic atoms, the phase model is considered 
a reasonable starting point to then get beyond-mean-field quantum effects 
\cite{smerzi2000,anglin2001,ferrini2008,sala2021}.

In this paper, we study a Josephson junction by using the Feynman path-integral 
approach \cite{nagaosa,wen}. In particular, we consider a system of 
interacting bosons which are tunneling between two sites. 
At the mean-field (saddle-point) level we recover the classical 
phase-imbalance model \cite{smerzi1997}. 
Performing path integration over the population imbalance 
we obtain the only-phase effective action of the system. 
The quantum effective only-phase action, which formally comprises 
all the quantum corrections, is then examined. 
The quantum correction to the Josephson frequency of oscillation 
is obtained to the second order in the derivative expansion and 
to the lowest order in $\hbar$. Finally, by using a 
different strategy based on the quantum average of the equation 
of motion, the same quantum correction is recovered. 
We also discuss the possible experimental detection of this 
quantum correction with atomic or superconducting 
Josephson junctions. 

\section{Two-site model}

The macroscopic quantum tunneling of bosonic particles or Cooper 
pairs in a Josephson junction made of two superfluids or two 
superconductors separated by a potential barrier can be described 
within a quantum field theory formalism. 

The simplest Lagrangian of a system made of bosonic particles 
which are tunneling between two sites ($j=1,2$) is given by 
\beq 
L = \sum_{j=1,2} \left[
\im \hbar \, \psi_j^* {\dot \psi_j} - {U\over 2} |\psi_j|^4 \right] 
+ {J\over 2} \left( \psi_1^*\psi_2 + \psi_2^*\psi_1 \right) ,
\eeq
where $\psi_j(t)$ is the adimensional complex field of bosons in the $j$ site 
at real time $t$, $U$ is the on-site interaction strength of particles, 
$J$ is the tunneling energy, $\hbar$ is the reduced Planck constant, 
$\im$ is the imaginary unit, and dot means the derivative 
with respect to time $t$. 

To make clear the crucial role of the hopping term, which contains 
the tunneling energy $J$, we set 
\beq 
\psi_j(t) = \sqrt{N_j(t)} \, e^{\im \phi_j(t)} \; , 
\label{madelung}
\eeq
where $N_j(t)$ is the number of the bosons in the 
$j$ site and $\phi_j(t)$ is the phase angle. 
We also introduce the total number 
\beq
N = N_1(t) + N_2(t) \;,
\label{ntotal}
\eeq
which is a constant of motion, and the relative phase 
\beq
\phi(t) = \phi_2(t) - \phi_1(t) , 
\label{phirelative}
\eeq
which is not a constant of motion, similarly to the total phase 
\beq 
{\bar \phi}(t) = \phi_1(t) + \phi_2(t) \; . 
\eeq 
We can then define the population imbalance as 
\beq
z(t) = { N_1(t) - N_2(t) \over N } \; .  
\label{popimb}
\eeq
In this way the Lagrangian becomes 
\beq 
\bal
&\bar{L}(\phi,z,\bar{\phi},N) = \frac{\im\hbar}{2}\dot{N}-\frac{U}{4}N^{2} \\
&+{N \hbar\over 2} \left( z {\dot \phi} - {\dot {\bar \phi}} 
\right) - {U N^2\over 4} z^2  
+ {J N\over 2} \sqrt{1-z^2} \, \cos{(\phi)} \; . 
\eal
\label{golden}
\eeq
The last term in this Lagrangian is the one that makes possible 
the periodic oscillation of a macroscopic number of particles 
between the two sites. Note that the constant term $-UN^2/4$ 
and the term $(N\hbar/2) {\dot {\bar \phi}}$ 
containing the exact differential ${\dot {\bar \phi}}$ 
can be safely removed. 

\section{Mean-field dynamics} 

The quantum mechanics of the Josephson junction can be derived 
from the Feynman path integral \cite{nagaosa,wen}
\beq 
\bal
&\int {\cal D}[\psi_{1}(t)]  {\cal D}[\psi_{1}^{*}(t)]
{\cal D}[\psi_{2}(t)]  {\cal D}[\psi_{2}^{*}(t)] 
\ \e^{\im \int {\bar L}(\phi,z,\bar{\phi},N)\, dt/\hbar} \\
&=\int {\cal D}[\phi(t)]  {\cal D}[z(t)]\int\mathcal{D}[N(t)]
\e^{\im \int 
\left[{\bar L}(\phi,z,\bar{\phi},N)-\hbar \dot{N}\bar{\phi}/2\right] 
dt/\hbar} \\
&\times\underbrace{\int\mathcal{D}[\bar{\phi}(t)]\e^{\im\int \dot{N}
\bar{\phi}dt/2}}_{2\delta[\dot{N}]} \\
&=\int {\cal D}[\phi(t)]  {\cal D}[z(t)] 
\ \e^{\im\int L(\phi,z) \, dt/\hbar} \; , 
\eal
\eeq
integrating over all the configurations of the dynamical 
variables $\phi(t)$ and $z(t)$ with
\beq
L(\phi,z)=\frac{N\hbar}{2}z\dot{\phi}-\frac{UN^{2}}{4}z^{2}+\frac{JN}{2}
\sqrt{1-z^{2}}\cos{(\phi)} \; , 
\label{gold}
\eeq
where $N(t)=N$ is a constant as a consequence of integrating out $\bar{\phi}$. 
Here, we have omitted a constant originating from the Jacobian for 
the Madelung transformation in Eq.~\eqref{madelung} and the transformation 
into relative coordinates given in Eqs.~\eqref{phirelative} and \eqref{popimb}: 
\beq
\bal
\mathcal{D}[\psi_{1}]\mathcal{D}[\psi_{1}^{*}]\mathcal{D}[\psi_{2}]
\mathcal{D}[\psi_{2}^{*}]
&=\mathcal{D}[N_{1}]\mathcal{D}[\phi_{1}]\mathcal{D}[N_{2}]
\mathcal{D}[\phi_{2}] \\
&=N\mathcal{D}[z]\mathcal{D}[\phi]\mathcal{D}[N]\mathcal{D}[\bar{\phi}] .
\eal
\eeq
Unfortunately, the exact calculation of these path integrals is an extremely difficult task also numerically, and consequently 
some approximation scheme is needed. 

The simplest approximation scheme to treat our quantum problem 
is the so-called mean-field (or saddle-point) approximation \cite{nagaosa,wen}, 
where one takes into account only the configurations which 
extremize the action functional 
\beq 
S = \int L(\phi,z) \, dt \; . 
\eeq
These configurations are the ones which satisfy the 
Euler-Lagrange equations. In our case, they are 
\beqa 
{\dot \phi} &=& J {z\over \sqrt{1-z^2}} \cos{(\phi)} + U N z \; ,
\label{starting1}
\\
{\dot z} &=& - J \sqrt{1-z^2} \sin{(\phi)} \; .  
\label{starting2}
\eeqa
These equations describe the mean-field dynamics of the macroscopic 
quantum tunneling in a Josephson junction, 
where $\phi(t)$ is the relative phase angle of the complex field of the superfluid (or superconductor) 
between the two junctions at time $t$ 
and $z(t)$ is the corresponding relative population imbalance 
of the Bose condensed particles (or Cooper pairs). 

It is important to stress that, due to the term 
$(N\hbar/2)z\dot{\phi}$ in the Lagrangian \eqref{gold}, the dynamical 
variables  $\phi(t)$ and $z(t)$ are canonically conjugated. 
This means that one can introduce the new dynamical variable 
\beq 
p_{\phi}(t) = {N\hbar\over 2} z(t) ,
\eeq
which is the generalized momentum conjugated to the Lagrangian 
coordinate $\phi(t)$. Moreover, with the Legendre transformation 
$H =p_{\phi} {\dot\phi} - L$, one obtains the Hamiltonian \cite{smerzi1997}
\beq 
H(\phi,p_{\phi})= {U p_{\phi}^2\over \hbar^2} - \frac{JN}{2} 
\sqrt{1-{4p_{\phi}^2\over N^2\hbar^2}}\ \cos{(\phi)} \;,
\label{diamond}
\eeq
of a nonrigid pendulum \cite{smerzi1997}. 
The Hamilton's equations of motion obtained with $H(\phi,p_{\phi})$ 
are merely Eqs.~\eqref{starting1} and \eqref{starting2}. 

\subsection{Linearized equations and Josephson oscillation}

Assuming that both $\phi(t)$ and $z(t)$ are small, 
i.e., $|\phi(t)|\ll 1$ and $|z(t)|\ll 1$, 
the Lagrangian \eqref{gold} can be approximated as 
\beq 
L^{(2)} = {N \hbar\over 2} z {\dot \phi} - 
{JN\over 4} \phi^2 - {(JN + U N^2)\over 4} z^2 \; ,  
\label{quadratic}
\eeq 
removing a constant term. The Euler-Lagrange equations of this 
quadratic Lagrangian are the linearized Josephson junction equations
\beqa 
\hbar \, \dot{\phi} &=& (J+UN) z  \; , 
\label{linear-jo1}
\\
\hbar \, \dot{z} &=& - J\phi \; ,  
\label{linear-jo2}
\eeqa
which can be rewritten as a single equation for the 
harmonic oscillation of $\phi(t)$ and the harmonic 
oscillation of $z(t)$, given by 
\beqa 
\ddot{\phi} + \Omega^2 \ \phi = 0 \; ,
\\
\ddot{z} + \Omega^2 \ z = 0 \; , 
\eeqa
both with frequency 
\beq 
\Omega = {1\over \hbar} \sqrt{J^2 + N U J} \; , 
\label{frequenza-vera-mf}
\eeq
that is, the familiar mean-field frequency of macroscopic quantum 
oscillation in terms of tunneling energy $J$, interaction strength $U$, 
and number $N$ of particles \cite{smerzi1997}. In the regime $NU/J \ll 1$ the 
frequency $\Omega$ becomes the Rabi frequency 
\beq 
\Omega_{\R}={J\over \hbar} \; , 
\eeq
while in the regime $NU/J \gg 1$ the frequency $\Omega$ 
becomes the Josephson frequency 
\beq 
\Omega_{\J}={\sqrt{NUJ}\over \hbar} \; . 
\label{jo-freq}
\eeq

\section{Effective only-phase action}

Fixing the initial and final
points for the $\phi(t)$ paths, while still summing over
all $z(t)$ paths, yields the path-integral propagator for the phase,
\begin{equation}
    K(\phi_T,T|\phi_0,0) =
    \int \limits_{ \{\phi_0,0\} }^{ \{ \phi_T,T \}  } {\cal D}[\phi]
    \int {\cal D}[z]\, \e^{\im S[\phi,z]/\hbar}.
\end{equation}
The quantum-mechanical way to derive an 
effective only-phase action $S_0$ for $\phi(t)$,
starting from the full action $S[\phi(t),z(t)]$, is to 
trace out the dynamical variable $z(t)$ with a path integral 
over it \cite{nagaosa,wen}, namely 
\begin{equation}
    \int {\cal D}[z]\, \e^{\im S[\phi,z]/\hbar}
    \propto \e^{\im S_0[\phi]/\hbar}.
\end{equation}
In our case, the action functional $S$ is determined by the
Lagrangian $L$ given by expression \eqref{golden}, containing 
both $\phi(t)$ and $z(t)$. 
However, the complete Lagrangian $L$ of Eq.~\eqref{golden} 
cannot be used to extract analytically this effective only-phase 
Lagrangian because one can explicitly calculate 
only quadratic integrals. To perform these calculations 
we can use quadratic expansions, i.e., the Gaussian approximation 
\cite{nagaosa,wen}.  
Expanding the Lagrangian $L$ of Eq.~\eqref{golden} at the Gaussian 
level with respect to $z(t)$ we obtain 
\beq 
L(\phi,z) = 
{N \hbar\over 2} z {\dot \phi} -{UN^2+JN\cos{(\phi)}\over 4} 
z^2 + {J N\over 2} \cos{(\phi)} \; . 
\label{Lphiz}
\eeq

To perform the integration over the $z(t)$ paths, it is useful
to use the time-sliced representation of the propagator. In that
case the paths are subdivided into $n$ time steps, chosen of
equal duration $\delta t=T/n$, and at the end of the calculation
one lets $n$ tend to infinity. The path integral over $z(t)$
is then performed as an $n$-fold integral over the variables
$z_j$, with $j=1,\cdots,n$. After performing these integrations,
we find that the propagator for the phase can be written as
\begin{equation}
    K(\phi_T,T|\phi_0,0) = 
    \left( \prod \limits_{j=1}^{n-1} \int \limits_{0}^{2 \pi} d\phi_j \right) 
    \prod \limits_{j=1}^{n} K_\textrm{inf} ( \phi_j,t_j | \phi_{j-1},t_{j-1}),
\end{equation}
where the infinitesimal propagator is given by
\begin{align}
    K_\textrm{inf} & ( \phi_j,\delta t | \phi_{j-1},0) = 
    \sqrt{\frac{N \hbar}{( U N + J \cos{\phi_j} ) 
4 \pi \im \delta t }}   \nonumber \\
    & \times \exp \left\{ -\frac{N \hbar (\phi_j-\phi_{j-1})^2 }
{4 \im ( U N + J \cos{\phi_j} ) \delta t} 
    \right. \nonumber \\
    &+ \left. \frac{\im}{2 \hbar} J N \cos{(\phi_j)} 
\delta t \right\}. \label{infiprop}
\end{align}
The prefactor before the exponential is a normalization factor that ensures 
the condition 
\begin{equation}
    \lim_{\delta t \rightarrow 0} K_\textrm{inf} ( \phi_j,\delta t | \phi_{j-1},0) 
= \delta(\phi_j-\phi_{j-1}).
\end{equation}
This prefactor is crucial to remove the divergences resulting from the 
quantum fluctuations of $\phi(t)$ 
\cite{bastianelli}. Letting the number $n$ of time slices go to infinity, 
the exponential phase factor 
for the $\phi$ path tends to
$\exp\left[\im S_0[\phi(t)]/\hbar\right]$ with 
\begin{equation}
    S_{0}[\phi]=\int dt 
    \left( \frac{N\hbar^{2}\dot{\phi}^{2}}{4\left(UN+J\cos{\phi}\right)}
    +{JN\over 2}\cos{\phi} \right) \; .
    \label{S0}
\end{equation}
The full propagator can then be written as
\begin{equation}
    K(\phi_T,T|\phi_0,0) = \int \limits_{ \{\phi_0,0\} }^{ \{ \phi_T,T \}  }
    {\cal D}[\phi] \e^{\im S_0[\phi]/\hbar},
\end{equation}
where now the path-integral measure is determined by the prefactor 
in Eq.~\eqref{infiprop}, 
and given by
\begin{equation}
    \int {\cal D}[\phi] = \lim_{n \rightarrow \infty} \prod \limits_{j=1}^n
      \int \limits_{-\pi}^{\pi} \frac{d\phi_j}
     { \sqrt{( U + J \cos{\phi_j}/N ) 4 \pi \im \delta t /\hbar} } . 
\label{measure}
\end{equation}

Note that the only-phase action $S_0[\phi]$ of Eq.~\eqref{S0} can
also be obtained substituting $z$ in the Lagrangian $L(\phi,z)$
of Eq.~\eqref{Lphiz} with the expression
\beq 
z = {\hbar \, {\dot \phi}\over UN + J \cos(\phi)} \; , 
\label{z-azzop}
\eeq
which is the Euler-Lagrange equation of $L(\phi,z)$ for
the dynamical variable $z(t)$. 
The Euler-Lagrange equation of the relative phase $\phi(t)$ derived from 
the action $S_0[\phi]$ of Eq.~\eqref{S0} is 
\beq 
{\hbar^2 {\ddot\phi}\over UN+J\cos(\phi)} 
+ {J\over 2} {\hbar^2 {\dot\phi}^2 \sin(\phi)\over (UN+J\cos(\phi))^2} 
+ J \sin(\phi) = 0 \; . 
\label{lippo-lillo}
\eeq
Its linearized version is 
\beq 
{\hbar^2\over UN+J}{\ddot\phi} + J \, \phi = 0 \; , 
\eeq
which gives again Eq.~\eqref{frequenza-vera-mf} for the Josephson frequency.

\subsection{Schr\"odinger equation for phase wavefunction} 

The Lagrangian of the action \eqref{S0} is given by 
\beq 
L_0 = 
\frac{N\hbar^{2}\dot{\phi}^{2}}{4\left[UN+J\cos{(\phi)}\right]}
+{JN\over 2}\cos{(\phi)}  \; , 
\eeq
and the corresponding generalized momentum $p_{\phi}$ reads 
\beq 
p_{\phi} = {\partial L_0\over \partial \dot\phi} = 
\frac{N\hbar^{2}\dot{\phi}}{2\left[UN+J\cos{(\phi)}\right]} \; . 
\eeq
As expected, $p_{\phi}$ is proportional to the population imbalance $z$
given by Eq.~\eqref{z-azzop}: $p_{\phi}=(N\hbar/2)z$. 
The Legendre transformation $H_0 = p_{\phi}{\dot\phi} - L_0$ 
gives the Hamiltonian 
\beq 
H_0(\phi,p_{\phi}) = 
\left[UN+J\cos{(\phi)}\right]{p_{\phi}^2\over N\hbar^2} - 
{JN\over 2}\cos{(\phi)} \; . 
\eeq
Clearly, this Hamiltonian can be obtained from the one 
of Eq.~\eqref{diamond} expanding the square root up to the quadratic 
term with respect to $p_{\phi}$. 

Promoting $p_{\phi}$ to the operator ${\hat p}_{\phi}=-\im\hbar \partial_{\phi}$ 
we can immediately write the time-dependent 
Schr\"odinger equation $\im\hbar\partial_t\Psi={\hat H}_0\Psi$, namely 
\beq 
\bal
&\im\hbar \partial_t \Psi(\phi,t) \\
&= 
\left[ - \left[U+{J\over N}\cos(\phi)\right] \partial_{\phi}^2 
- {JN\over 2}\cos{(\phi)} \right] \Psi(\phi,t) \; , 
\eal
\eeq
for the wave function $\Psi(\phi,t)$ of the relative phase $\phi$. 
Moreover, the quantum Hamiltonian 
${\hat H}_0=H_0({\hat \phi},{\hat p}_{\phi})$ is such that 
\beqa 
\langle \phi_b| \e^{-\im{\hat H}_0 (t_b-t_a)/\hbar}|\phi_a 
\rangle &=& \int \mathcal{D}[\phi] \mathcal{D}[p_{\phi}] \e^{{\im\over\hbar} 
\int_{t_a}^{t_b} \left[ {\dot\phi}  p_{\phi} - H_0(\phi,p_{\phi}) \right]dt} 
\nonumber 
\\
&=& \int \mathcal{D}[\phi]\, \e^{\im S_{0}[\phi]/\hbar} \; . 
\eeqa 
Thus, we recover the effective action $S_0[\phi]$ 
of Eq.~\eqref{S0} after functional 
integration over the generalized momentum $p_{\phi}$. 
The integration measure \eqref{measure} arises here from the presence of an unusual 
kinetic term in the Hamiltonian $H_0(\phi,p_{\phi})$ \cite{abers,ryder,bastianelli}. 
 
\section{Quantum effective only-phase action}
 
It is important to stress that under condition $UN\gg J$ (Josephson regime), the integration measure \eqref{measure} does not depend on $\phi(t)$. 
Moreover, the action $S_0[\phi]$ becomes 
\beq
S_{\J}[\phi]= \int dt \left[\frac{\hbar^{2}\dot{\phi}^{2}}{4U}+{JN\over 2}\cos({\phi})\right] \; . 
\label{resam}
\eeq
In this regime, we can ignore a quartic contribution of $\phi$ in the kinetic term in Eq.~\eqref{S0}. 
Equation \eqref{resam} is merely the familiar only-phase action of a 
capacitively shunted superconducting 
Josephson junction, where the population imbalance is very small, 
i.e., $|z(t)|\ll 1$, but the number of bosonlike 
Cooper pairs $N$ is large. 
Within the framework of superconducting junctions, $I_0=eJN/\hbar$ 
is the critical electric current with $e$ the electric charge 
of the electron, and $C=2e^2/U$ is the electric capacitance \cite{nagaosa}. 

Let us work with the action of Eq.~\eqref{resam}. We want to 
determine the corrections to the Josephson frequency 
due to quantum fluctuations adopting the formalism of the quantum 
effective action \cite{goldstone,jona1,coleman}. The quantum effective 
action $\Gamma[\phi]$ is a modified expression of the action $S_\J[\phi]$ 
which takes into account quantum corrections. 
The minimization of $\Gamma[\phi]$ gives the exact equations of motion for 
the expectation value of the field, which is here denoted 
with the same symbol $\phi(t)$ of the field \cite{goldstone,jona1,coleman}. 
First, we rewrite the action \eqref{resam} as follows, 
\beq 
S_{\J}[\phi]= \int dt \left[ {M_\J\over 2} \dot{\phi}^2 
- V(\phi) \right] \; , 
\label{resam-l}
\eeq
where $M_\J=\hbar^2/(2U)$ and the only-phase potential energy is 
expanded as 
\beq 
V(\phi) = \frac{M_\J \Omega_{\J}^2}{2} \phi^2 + {\tilde V}(\phi) ,
\eeq
with $\Omega_{\J}$ given by Eq.~\eqref{jo-freq} and 
\beq
\bal 
{\tilde V}(\phi)
&=M_{\J}\Omega_{\J}^{2}\left[1-\cos({\phi})\right]-
\frac{M_{\J}\Omega_{\J}^{2}}{2}\phi^{2} \\
&= {\lambda} \, \phi^4 + O(\phi^6) \; ,
\eal
\label{u-phi}
\eeq 
with $\lambda=-JN/48$. 
As discussed in Ref.~\cite{jona2}, it is the potential ${\tilde V}(\phi)$ which encodes quantum fluctuations within the formalism of the quantum effective action, where the action $S_{\J}[\phi]$ is substituted by the quantum effective action $\Gamma_{\J}[\phi]$ given by 
\beq 
\Gamma_{\J}[\phi]= \int dt \left[ {Z(\phi)\over 2} \dot{\phi}^2 
- V_{\text{e}}(\phi) \right] ,
\eeq
to the second order in the derivative expansion \cite{coleman,jona1,jona2}. 
In this quantum effective action, the term 
\beq 
Z(\phi) = M_{\J} + \hbar \, Z_1(\phi) + \hbar^2 Z_2(\phi) + \cdots ,
\eeq
is the effective mass and 
\beq 
V_{\text{e}}(\phi) = V(\phi) + \hbar \, V_{\text{e}1}(\phi) + \hbar^2 \, V_{\text{e}2}(\phi) + \cdots , 
\eeq
is the effective potential, written in terms of a $\hbar$ expansion 
\cite{jona2}. In particular, to the first order of this $\hbar$ expansion 
one finds \cite{jona2}
\beq 
Z_1(\phi) = {1\over 32 M_J^2} 
{\left[ \partial_{\phi}^3{\tilde V}(\phi) \right]^2 \over 
\left[\Omega_{J}^2 + \dfrac{1}{M_\J} \partial_{\phi}^2{\tilde V}
(\phi)\right]^{5/2}} ,
\label{Z1}
\eeq
and 
\beq 
V_{\text{e}1}(\phi) = {1\over 2} \left( \sqrt{\Omega_{\J}^2 + 
{1\over M_\J} \partial_{\phi}^2 {\tilde V}(\phi)} - \Omega_{\J}\right) \; . 
\label{Ve1}
\eeq
Note that the potential $V(\phi)=M_{\J}\Omega_{\J}^{2}/2+\Tilde{V}(\phi)$ must be convex so that Eqs.~\eqref{Z1} and \eqref{Ve1} are real. 
This is realized in the domain $\abs{\phi}<\sqrt{2}$, which has already been satisfied under $\abs{\phi}\ll1$ that makes the approximation from Eq.~\eqref{S0} to Eq.~\eqref{resam} valid. 
The corresponding equation of motion with first-order quantum corrections, 
obtained extremizing $\Gamma_\J[\phi]$, is given by \cite{jona2}
\beq 
\left[ M_{\J} + \hbar Z_1(\phi)\right] {\ddot \phi} + 
{\hbar\over 2} \partial_{\phi}Z_1(\phi) \, {\dot\phi}^2 = 
- {\partial_{\phi}} \left[ V(\phi) + \hbar \, V_{\text{e}1}(\phi) \right] \; . 
\label{EoMeff1}
\eeq
The equation of motion \eqref{EoMeff1} is derived through the second-order derivative expansion and neglecting higher-order contributions of $\hbar$. 
This approximation is valid under \cite{jona2}
\beq
\abs{\frac{\lambda\phi^{4}}{M_{\J}\Omega_{\J}^{2}\phi^{2}/2}}\ll1 ,
\label{ineq1}
\eeq
and, by inserting the oscillator length $\phi=\sqrt{\hbar/\left(M_{\J}\Omega_{\J}\right)}$, 
\beq
\abs{\frac{2\hbar\lambda}{M_{\J}^{2}\Omega_{\J}^{3}}}\ll1 .
\label{ineq2}
\eeq
Inequality \eqref{ineq1} reads $\abs{\phi}\ll2\sqrt{3}$, which should have already been satisfied under $\abs{\phi}\ll1$. 
The latter one \eqref{ineq2} reads $UN/J\ll 36N^{2}$. 
This inequality indicates that the approximation is valid under a sufficiently small ratio between the interaction energy and the tunneling energy within the Josephson regime $1 \ll UN/J \ll 36N^{2}$. 

Because we want to determine the first-order quantum 
correction to the Josephson frequency, we consider $\Tilde{V}(\phi)$ 
of Eq.~\eqref{u-phi} up to the quadratic term of $\phi$. 
In this way, up to the quadratic term of $\phi$, we obtain 
\beq
Z_1(\phi) = 0 ,
\eeq
and 
\beq 
\bal
V_{\text{e}1}(\phi) 
&=\frac{\Omega_{\J}}{2}\left(\sqrt{\cos{\phi}}-1\right) \\
&\simeq {3\lambda\over M_\J \Omega_{\J}}\, \phi^2 
= - {\sqrt{NUJ}\over 8\hbar} \, \phi^2 \; .
\eal
\eeq
It follows that, to the first 
order of $\hbar$, the modified frequency of oscillation is given by 
\beq 
{\tilde \Omega}_{\J} = {\sqrt{NUJ}\over \hbar} \sqrt{ 1 - {1\over 2} 
\sqrt{{U\over JN}}} \; .
\label{magicabula}
\eeq
Clearly, for $N\gg 1$ one recovers Eq.~\eqref{jo-freq}. 
In the framework of the the capacitively-shunted superconducting 
Josephson junction, where $U=2e^2/C$ and $JN=(\hbar/e)I_0$, 
the frequency can be written as 
\beq 
{\tilde \Omega}_{\J} = {1\over \hbar} \sqrt{2 e\hbar I_0\over C} 
\sqrt{ 1 - {1\over 2} \sqrt{2e^3\over \hbar C I_0}} \; .  
\label{magicabula2}
\eeq
Reference \cite{pigneur} reports the Josephson dynamics 
with two one-dimensional quasicondensates of ${}^{87}\mathrm{Rb}$ atoms 
trapped in a double-well potential. 
The number of atoms is typically $N=2500$ and the ratio between the 
interaction energy and the tunneling energy is $NU/(2J)\sim10^{2}$ 
\cite{pigneur}. 
These experimental data give, for the relative difference between the mean-field Josephson frequency $\Omega_{\J}$ and the beyond-mean-field one ${\tilde \Omega}_{\J}$, the value $(\Omega_{\J}-\Tilde{\Omega}_{\J})/\Omega_{\J}\simeq 0.1\%$, which indicates that the quantum correction slightly reduces the Josephson frequency. 
One could observe the effect of this quantum correction more 
significantly in the deep Josephson regime by increasing the ratio 
$NU/J$ at fixed $N$, namely, by 
increasing the interparticle interaction strength $U$ or decreasing the 
Josephson coupling $J$ while satisfying Eq.~\eqref{ineq2} because $\Tilde{\Omega}_{\J}/\Omega_{\J}=
[1-\Omega_{\J}/(2N\Omega_{\R})]^{1/2}$ with $\Omega_{\J}/\Omega_{\R}=\sqrt{NU/J}$. 
Considering, instead, superconducting Josephson junctions, 
in Ref.~\cite{devoret} the experimental value 
$2e\Omega_{\J}/I_{0}=2[2e^{3}/(\hbar CI_{0})]^{1/2}\simeq2.3\times10^{-3}$ 
results in $(\Omega_{\J}-\Tilde{\Omega}_{\J})/\Omega_{\J}\simeq0.03\%$, which reduces the Josephson frequency due to the quantum correction in the order 
of $10^{-4}$. 
A larger value of $2e\Omega_{\J}/I_{0}$ would make the quantum correction more significant.
 
\section*{Alternative derivation of quantum corrections}

In this section, we discuss a different approach to derive 
the quantum correction of the only-phase dynamics. 
This approach is similar to the one recently developed to determine 
beyond-mean-field corrections to the critical temperature 
in two-band superconductors \cite{sala2019}. 
Given the nonlinear only-phase Josephson equation 
\beq 
{\hbar^2\over UN}{\ddot\phi} + J \, \sin(\phi) = 0 \;,
\eeq
derived from Eq.~\eqref{resam}, we set
\beq 
\phi(t) = \phi_0(t) + {\tilde \phi}(t) \; , 
\eeq
where $\phi_0(t)$ is the mean-field solution and ${\tilde \phi}(t)$ encodes
quantum fluctuations, which are assumed to be small. At
the quadratic level with respect to ${\tilde \phi}(t)$ we get 
\beqa 
{\hbar^2\over UN}{\ddot\phi}_0 &+&
{\hbar^2\over UN}{\ddot{\tilde \phi}} + J \, \sin(\phi_0)
\nonumber 
\\
&+& J \, \cos(\phi_0) \, {\tilde \phi} 
- {J\over 2} \, \sin(\phi_0) \, {\tilde \phi}^2 = 0 \; . 
\eeqa
Performing the quantum average $\langle\cdots\rangle $ with the condition 
$\langle \phi(t)\rangle = \phi_0(t)$, i.e., $\langle {\tilde\phi}(t)\rangle 
= 0$, we obtain
\beq
{\hbar^2\over UN}{\ddot\phi}_0 + J \left( 1 - {1\over 2} 
\langle {{\tilde \phi}^2} \rangle\right) \, \sin(\phi_0) = 0 \; .
\eeq
The quantum average $\langle {{\tilde \phi}^2} \rangle$ 
can be calculated as follows, 
\beq 
\langle  {\tilde\phi}^2\rangle = {1\over {\cal Z}}
\int D[{\tilde \phi}] \, {\tilde\phi}^2 \, \e^{\im S^{(2)}_\J[{\tilde\phi}]/\hbar} \; , 
\eeq
where 
\beq 
S_\J^{(2)}[{\tilde \phi}] = \int \left[{\hbar^2\over 4U} 
{\dot{\tilde\phi}}^2 - {JN\over 4} {\tilde\phi}^2 \right] \ dt
\eeq
is the quadratic action for the fluctuations and 
\beq 
{\cal Z} = \int D[{\tilde \phi}] \, 
\e^{\im S^{(2)}_\J[{\tilde\phi}]/\hbar}  
\eeq
is the corresponding real-time partition function. 
Then, one easily finds the zero-temperature result
\beq 
\langle {\tilde \phi}^2\rangle = \sqrt{{U\over JN}} \; .
\eeq
Thus, the only-phase Josephson equation corrected by
quantum fluctuations reads
\beq 
{\hbar^2\over UN} {\ddot \phi}_0 + J \left( 1 - {1\over 2} 
\sqrt{{U\over JN}}\right) \, \sin(\phi_0) = 0 \; .
\eeq
It follows that the frequency of oscillation 
with the inclusion of quantum fluctuations is 
given by Eq.~\eqref{magicabula}. Thus, we have recovered 
the same result obtained with the quantum effective 
action formalism. 

\section{Discussion and conclusions}

In this paper, we have adopted a quantum field theory formalism based on the path integral to study the role of quantum fluctuations in a Josephson junction. 
From the bosonic action of the relative phase and population imbalance,  
by performing Gaussian integration over the population 
imbalance, under the condition of taking only up to quadratic terms in 
the population imbalance (i.e., with an approximated phase-imbalance action), 
we have derived the effective only-phase action. 
Quite remarkably, this effective action is highly nonlinear 
with respect to the phase variable but it gives rise to the same 
mean-field equation of the approximated phase-imbalance action. 
We have then examined the quantum effective only-phase action, 
which formally comprises all the quantum corrections for the 
dynamics of the expectation value of the relative phase. 
In this way, we have obtained, with two independent but similar procedures, 
the quantum-corrected Josephson frequency ${\tilde\Omega}_{\J}$, 
Eqs.~\eqref{magicabula} and \eqref{magicabula2}. 
As we have discussed in the last section, the estimated quantum corrections to the Josephson frequencies in a Bose-Josephson junction and in a superconducting Josephson junction are relatively small based on the current experiments \cite{pigneur,devoret}. 
Reference \cite{barankov} is an earlier work that has also considered quantum corrections to the Josephson effects. 
They started from a spatially three-dimensional Bose gas to obtain the correction to the Josephson energy. 
A different point between Ref.~\cite{barankov} and our work is that we are treating a spatially zero-dimensional system for simplicity. 
In Ref.~\cite{barankov}, instead, they integrated out the noncondensed field after separating the field operators into a condensed field and a noncondensed field following the Bogoliubov prescription. 
This approach has led to corrections to the Josephson energy originating from the interparticle interaction and the temperature in Bose condensates. 
Our result of modified Josephson frequencies, however, can be verified not only in an atomic Josephson junction but also in a superconducting Josephson circuit. 
We expect that tuning the experimental parameters such as the interaction strength or capacitance enables us to observe more prominent quantum corrections.

\section*{Acknowledgements}

We acknowledge Alessio Notari for discussions. 
L.S. thanks Fiorenzo Bastianelli, Alberto Cappellaro, Andrea Tononi, and Carlo Presilla for useful discussions. 
K.F. is supported by a Ph.D. fellowship of the Fondazione Cassa di Risparmio di Padova e Rovigo.


\begin{thebibliography}{999}

\bibitem{josephson1962} B. D. Josephson, Phys. Lett. \textbf{1}, 251 (1962).

\bibitem{barone1982} A. Barone and G. Paterno, 
\textit{Physics and Applications of the Josephson effect} 
(Wiley, New York, 1982). 

\bibitem{vari2017} E.L. Wolf, G.B. Arnold, M.A. Gurvitch, and 
John F. Zasadzinski, \textit{Josephson Junctions: History, Devices, 
and Applications} (Pan Stanford Publishing, Singapore, 2017). 

\bibitem{smerzi1997} A. Smerzi, S. Fantoni, S. Giovanazzi, and S.R. Shenoy, 
Phys. Rev. Lett. {\bf 79}, 4950 (1997). 

\bibitem{leggett1991} A. Leggett and F. Sols, Found. Phys. 
{\bf 21}, 353 (1991).

\bibitem{luis1993} A. Luis and L.L. Sanchez-Soto, 
Phys. Rev. A {\bf 48}, 4702 (1993). 

\bibitem{smerzi2000} A. Smerzi and S. Raghavan, 
Phys. Rev. A \textbf{61}, 063601 (2000).

\bibitem{anglin2001} J.R. Anglin, P. Drummond, and A. Smerzi, 
Phys. Rev. A {\bf 64}, 063605 (2001).
	
\bibitem{ferrini2008} G. Ferrini, A. Minguzzi, and F.W.J. Hekking,
Phys. Rev. A \textbf{78}, 023606 (2008).

\bibitem{sala2021} S. Wimberger, G. Manganelli, A. Brollo, and L. Salasnich, 
Phys. Rev. A {\bf 103}, 023326 (2021). 
	
\bibitem{nagaosa} N. Nagaosa, \textit{Quantum Field Theory in 
Condensed Matter Physics} (Springer, Berlin, 2013).
	
\bibitem{wen} X.-G. Wen, \textit{Quantum Field Theory of Many-Body 
Systems: from the Origin of Sound to an Origin of Light and Electrons},
(Oxford University Press, Oxford, U.K., 2004).

\bibitem{bastianelli} F. Bastianelli and P. van Nieuwenhuizen, 
{\it Path Integrals and Anomalies in Curved Space} 
(Cambridge University Press, Cambridge, U.K., 2006). 

\bibitem{abers} E.S. Abers and B.W. Lee, Phys. Rep. {\bf 9}, 1 (1973). 

\bibitem{ryder} L. Ryder, {\it Quantum Field Theory} 
(Cambridge University Press, Cambridge, U.K., 1996). 

\bibitem{goldstone} S. Weinberg and J. Goldstone, Phys. Rev. {\bf 127}, 
965 (1962). 

\bibitem{jona1} G. Jona-Lasinio, Nuovo Cimento {\bf 34}, 1790 (1964). 

\bibitem{coleman} S. Coleman and E. Weinberg, Phys. Rev. D. {\bf 7}, 
1888 (1973). 

\bibitem{jona2} F. Cametti, G. Jona-Lasinio, C. Presilla, 
and F. Toninelli, in {\it New Directions in Quantum Chaos}, Proceedings of International School of Physics ``Enrico Fermi'', 
Course CXLIII, Varenna, 1999, edited by G. Casati, I. Guarnieri, and U. Smilansky (IOS Press, Amsterdam, 2000), pp. 431-448. 

\bibitem{pigneur} M. Pigneur, T. Berrada, M. Bonneau, T. Schumm, E. Demler, 
and J. Schmiedmayer, Phys. Rev. Lett. {\bf 120}, 173601 (2018). 

\bibitem{devoret} M.H. Devoret, J.M. Martinis, and J. Clarke, 
Phys. Rev. Lett. {\bf 55}, 1908 (1985). 

\bibitem{sala2019} L. Salasnich, A.A. Shanenko, A. Vagov, J. Albino 
Aguiar, and A. Perali, Phys. Rev. B {\bf 100}, 064510 (2019). 

\bibitem{barankov} R.A. Barankov and S.N. Burmistrov, Phys. Rev. A {\bf 67}, 013611 (2003). 

\end{thebibliography}
\end{document}